# Disinformation and Social Bot Operations in the Run Up to the 2017 French Presidential Election

Emilio Ferrara
University of Southern California, Information Sciences Institute

## Abstract
Recent accounts from researchers, journalists, as well as federal investigators, reached a unanimous conclusion: social media are systematically exploited to manipulate and alter public opinion. Some disinformation campaigns have been coordinated by means of bots, social media accounts controlled by computer scripts that try to disguise themselves as legitimate human users. In this study, we describe one such operation occurred in the run up to the 2017 French presidential election. We collected a massive Twitter dataset of nearly 17 million posts occurred between April 27 and May 7, 2017 (Election Day). We then set to study the *MacronLeaks* disinformation campaign: By leveraging a mix of machine learning and cognitive behavioral modeling techniques, we separated humans from bots, and then studied the activities of the two groups taken independently, as well as their interplay. We provide a characterization of both the bots and the users who engaged with them, and oppose it to those users who didn't. Prior interests of disinformation adopters pinpoint to the reasons of scarce success of this campaign: the users who engaged with *MacronLeaks* are mostly foreigners with preexisting interest in alt-right topics and alternative news media, rather than French users with diverse political views. Concluding, anomalous account usage patterns suggest the possible existence of a black-market for reusable political disinformation bots.

## Introduction

Social media have been extensively praised for their power to democratize online conversation. Whether in the context of civil movements (Howard, et al., 2011; González-Bailón, et al., 2011; Tufekci & Wilson, 2012; González-Bailón, et al., 2013; Tufekci, 2014; Bastos, et al., 2014), political outreach (Bond, et al., 2012; Bakshy, et al., 2015), public health interventions (Centola, 2010; Dredze, 2012; Korda & Itani, 2013), or situational awareness (Sasaki, et al., 2010; Merchant, et al., 2011; Signorini, et al., 2011; Paul & Dredze, 2011), platforms like Twitter and Facebook play a central role in the modern information ecosystem. However, such powerful tools can also be abused for nefarious purposes (Ferrara, 2015): Extremist groups use social media for radical propaganda and recruitment efforts (Ferrara, 2017); stock market manipulators have created concerted efforts to game financial systems (Ferrara, 2015); conspiracy groups orchestrate campaigns to distribute fake scientific articles to support anti-vaccination and other anti-science movements, creating massive public health issues (Bessi, et al., 2015; Del Vicario, et al., 2016). Of great concern for democracy is another form of social media manipulation: The rise of popularity of bots and disinformation in the context of political propaganda (El-Khalili, 2013; Bessi & Ferrara, 2016; Ferrara, et al., 2016; Shorey & Howard, 2016; Kollanyi, et al., 2016; Marwick & Lewis, 2017).

Researchers warned about the potential for abuse of the social media ecosystem for political propaganda a decade ago (Howard, 2006; Hwang, et al, 2012). The earliest reports of coordinated attacks against political candidates on social media date back to 2010 (Metaxas & Mustafaraj, 2012; Ratkiewicz, et al., 2011a; Ratkiewicz, et al., 2011b). Since then, an increasing account of such events has been recorded in the context of several elections, both in the United States (Bessi & Ferrara, 2016; Kollanyi, et al., 2016; Shorey & Howard, 2016; Wolley, 2016; Wolley & Howard, 2016; Marwick & Lewis, 2017; Wang, et al., 2017), and all over the world, including in South America (Forelle, et al., 2015; Suárez-Serrato, et al., 2016), the U.K. (Howard & Kollanyi, 2016), and Italy (Cresci, et al., 2017). One common trait of these campaigns is the adoption of automation tools to generate large volume of social media posts to support, or attack, candidates.

Although automated social media accounts, referred to as *social bots* or *sock puppets*, have in some instances been used for social good (Savage, et al., 2016; Mønsted, et al., 2017; Shirado, & Christakis, 2017), in this study we will refer exclusively to those used with the intent to deceive and manipulate. Another form of artificial support expressed on social media is via fake followers, often inactive accounts that are only used to increase the online popularity and/or visibility (e.g., the followership) of a public figure (Marwick & Lewis, 2017). Fake followers are not subject of this study due to their inactivity.

On the onset of the 2016 U.S. presidential election, a rather new phenomenon was observed, in concert with social bots and hyper-partisan campaigns: the spread of fake news and the coordination of disinformation campaigns (Allcott & Gentzkow, 2017; Marwick & Lewis, 2017; Mele, et al., 2017). The adoption of automated devices such as social bots in the context of disinformation campaigns is particularly concerning because there is the potential to reach a critical mass large enough to dominate the public discourse and alter public opinion (Ferrara, et al., 2016; Woolley & Howard, 2016; Marwick & Lewis, 2017); this could steer the public's attention away from facts and redirecting it toward manufactured, planted information.[1]

In this paper, we focus on another pivotal recent political event, namely the 2017 French presidential election of May 7, 2017. We aim to describe the social media dynamics related to one potentially disruptive disinformation campaign that occurred in the run up to the election, known as "MacronLeaks". In the following, we provide a brief account of the events related to *MacronLeaks* as they unfolded.

The popular 4chan.org message board hosts several large yet ephemeral discussion threads. A popular board is the "/pol/" (i.e., politics) community. Partly due to its anonymity features, partly because inactive discussion threads are quickly and automatically archived by the platform itself, 4chan.org has been reportedly functioning as an effective incubator of alt-right and alt-right online communities, most prominently in the United States (Marwick & Lewis, 2017). Among the threads related to the 2017 French election, the most popular was one centered around coordinating cyber-attacks aimed at reveling sensitive information about then-presidential candidate Emmanuel Macron (En Marche!, 2017).

Significant user participation guaranteed the generation of a wealth of allegedly incriminating material: while most documents were manufactured and their false nature was easily identifiable,[2] on Friday May 5, 2017 one user anonymously shared an email dump containing "Correspondence, documents, and photos from Macron and his team."[3] Shortly after its appearance, a link to the "/pol/" anonymous post was shared on Twitter by alt-right activist Jack Posobiec, which contributed to amplify the ongoing disinformation campaign object of this study.[4] Ultimately, the leaked documents were shared on Twitter by WikiLeaks' official account itself, although with the disclosure that their authenticity was unverified – this made the campaign go viral.

## DEFINING DISINFORMATION IN THE CONTEXT OF THIS STUDY

The notion of "disinformation campaign" used in this work warrants a rigorous definition. We consider *MacronLeaks* an instance of disinformation as it exhibited two necessary ingredients namely, first, the **unverified nature** of the shared information, and second the **coordinated effort** behind its sharing.

The unverified nature of the leaked documents make this information qualify for the traditional definition of "rumor" (Allport & Postman, 1947). The documents circulated online have been referred to as evidence of Macron's tax frauds and other illicit activities. The accuracy and relevance of these leaked documents has been extensively debated for weeks after the fact. The conclusion from official investigations is that, although the leaked documents were not manufactured, there is no evidence to support the allegations: most of the conclusions drawn by the 4chan.org communities were based on erroneous translation from French, as well as biased interpretations (or trivial misunderstanding) of French law (En Marche!, 2017).

The voluntary spreading of a rumor by means of an orchestrated effort makes it a disinformation campaign (Del Vicario, et al., 2016). In this case, 4chan.org served as incubator for the initial attempt to smear the candidate by manufacturing, planting, or leaking allegedly incriminating documents. In addition, one major finding of this work, that we will discuss in detail later, is the uncovering of a social bot operation that occurred in the run up to Election Day aimed at amplifying even further the viral sharing of disinformation and email leaks.

## SUMMARY OF CONTRIBUTIONS

In the rest of this manuscript, we will detail the methodological approaches to data collection and analysis, and discuss the findings and contributions of this work, which we summarize as follows:

- We monitored the Twitter stream between April 27 and May 7, 2017 (Election Day), and collected a very large dataset containing nearly 17 million tweets related to the 2017 French presidential election. Within it, we identified the subset related to the *MacronLeaks* disinformation campaign.
- By exploiting combinations of machine learning techniques and cognitive behavioral modeling, we identified humans and social bots participating to the campaign. We studied the characteristics of both classes of users independently, as well as their interplay.
- We discovered that prior user interests reveal the reasons of scarce success of the MacronLeaks campaign: the users who engaged with it were mostly foreigners belonging to the alt-right Twitter community, rather than French users (i.e., potential voters).
- Finally, we uncovered that accounts used to support then-presidential candidate Trump before the 2016 U.S. election have been brought back from a limbo of inactivity (since November 2016) to join the MacronLeaks disinformation campaign. Such anomalous usage patterns point to the possible existence of a black-market for reusable political-disinformation bots.

# METHODS

## DATA COLLECTION

By following a consolidated strategy, which we previously used to study the 2016 U.S. Presidential election (Bessi and Ferrara, 2016), we manually selected a set of hashtags and keywords. The list by construction contains a roughly equal number of terms associated with each of the two candidates, namely Marine Le

Pen and Emmanuel Macron, and various general election-related terms: we finally defined 23 terms, listed in **Table 1**. The compilation of this list was also informed by early reports by Oxford researchers about the most prominent election-related terms (Desigaud, et al., 2017; Howard, et al., 2017).

We monitored the Twitter stream and collected data by using the Twitter Search API, from April 27 to the end of election day, on May 7, 2017: this allowed us to uninterruptedly collect all tweets containing any of the search terms. The data collection infrastructure ran inside USC servers to ensure resilience and scalability. We chose to use the Twitter Search API (https://dev.twitter.com/rest/public/search) to make sure that we obtained all tweets that contain the search terms of interest posted during the data collection period, rather than a sample of unfiltered tweets: this precaution avoids incurring in known sampling issues related to collecting data using the Twitter Stream API (https://dev.twitter.com/streaming/overview) rather than the Twitter Search API (Morstatter, et al., 2013).

This procedure yielded a large dataset containing approximately 17 million unique tweets, posted by 2,068,728 million unique users. The timeline of the volume of posted tweets, down to the resolution of minutes, is illustrated in **Figure 1** – and it will be discussed more thoroughly in the **Data Analysis** section.

Within this set, we further identified the subset of tweets associated with the *MacronLeaks* campaign. Using the reports from various news articles as well as by manual inspection, we used the following key terms to isolate tweets belonging to the campaign: #MacronLeaks, #MacronGate, #SortonsMacron, #Bayrougate, and #RejoignezMarine. All tweets containing any such terms constitute the *MacronLeaks* corpus that we will analyze in detail in this work. Nearly 350 thousand such tweets exist, out of about 17 million overall tweets, making the MacronLeaks campaign a small drop (i.e., about 2% of the total) in the ocean of information regarding the presidential election. Overall, 99,378 users posted tweets in the MacronLeaks corpus.

## BOT DETECTION TECHNIQUES

One of the most daunting tasks in social media analysis is determining whether a user account is controlled by a human or a software (i.e., a bot). A great deal of research aims to address this issue, including our own efforts (Ferrara, et al., 2016; Varol, et al., 2017) and others' (Messias, et al., 2013; Yang, et al., 2014; Gilani, et al., 2015; Subrahmanian, et al., 2016).

In this realm, *Botometer*[6] (formerly BotOrNot) represents, as of today, the only openly accessible solution (Davis, et al., 2016). It consists of an Application Programming Interface (API) developed in Python which allows to programmatically interact with the underlying machine learning system. *Botometer* has been proven quite accurate in detecting social bots (Davis, et al., 2016; Varol, et al., 2017),

However, the public interface of *Botometer* has two limitations that prevented us to use it in this project: the framework relies on the Twitter API to collect recent data about the accounts to inspect. The Twitter API imposes very strict query rate limits, therefore making it impossible to analyze more than a few thousand accounts with the public *Botometer* Python API. In this study, our goal is to detect bots in a very large population of over 2 million users, requiring an ad hoc large-scale bot detection solution. The second limitation is once again derived by the Twitter API: when *Botometer* inspects an account that has been either suspended, protected, quarantined, or deleted, the Twitter API does not provide any details about it, rendering *Botometer* unable to make any determination. Since this study will show that a significant portion of bot accounts involved in *MacronLeaks* has been either suspended, quarantined, or deleted shortly after Election Day (May 7, 2017), *Botometer* would not represent a suitable tool to analyze them.

## OUR BOT DETECTION APPROACH

For the reasons mentioned above, we decided to implement a simple yet accurate bot-detection algorithm reflecting the following requirements:

- The algorithm is accurate yet scalable and can be used to classify the over 2 million users present in our dataset in a reliable yet timely manner; this will address the scalability issues of *Botometer*.
- The algorithm can use historical tweets and account metadata collected and available in our dataset to determine bots and humans, without the need to query the Twitter API for recent data; this will address the limits imposed by the Twitter API and allow us to analyze all users, not only the active ones at the time of inspection, but also the suspended, protected, quarantined, and deleted accounts.
- The algorithm builds on top of the insights and lessons learned from the development of *Botometer*.

*Botometer*'s underlying machine-learning framework generates a set of over one thousand features, spanning content and network structure, temporal activity, user profile data, and sentiment analysis. These indicators are aggregated and analyzed to determine the likelihood that the inspected account is a bot. Feature analysis revealed that the two most important classes of feature to detect social bots are the metadata and usage statistics associated with a user account (Davis, et al., 2016; Varol, et al., 2017). We previously illustrated (Bessi and Ferrara, 2017) that the following indicators provide the strongest signals to separate humans from, in particular, political bots: *(i)* whether the public Twitter profile looks like the default one or it is customized (it requires human efforts to customize a profile, therefore bots are more likely to exhibit the default setting); *(ii)* absence of geographical metadata (humans often use smartphones and the Twitter iPhone/Android App, which records the physical location of the mobile device as digital footprint); and, *(iii)* activity statistics such as total number of tweets and frequency of posting (bots exhibit incessant activity and excessive amounts of tweets), proportion of retweets over original tweets (bots retweet contents much more frequently than generating new tweets), proportion of followers over followees (bots usually have less followers and more followees), the number of times a user has been added to a public list – human users are often considered more influential (Aral, & Walker, 2012) and their content more "contagious" (Kramer, et al., 2014; Ferrara & Yang, 2015a; Ferrara & Yang, 2015b; Mønsted, et al., 2017), etc.

## DETECTION ALGORITHM AND FEATURES

Considering these insights, we used the following user metadata and activity features to create a simple yet effective bot detection algorithm:[7]

1) "statuses_count": number of tweets posted by the given user;
2) "followers_count": number of followers of the given user;
3) "friends_count": number of followees (friends) of the given user;
4) "favourites_count": number of favorited tweets of the given user;
5) "listed_count": number of times the given user has been added to a list;
6) "default_profile": binary field that indicates whether the user profile has the default setting or not;
7) "geo_enabled": binary field that indicates whether the geo-coordinates of the user are available;
8) "profile_use_background_image": binary field that indicates whether the user profile has the default image or a custom one;
9) "verified": binary field that indicates whether the account has been verified by Twitter; verified accounts are considered to belong without doubt to humans.[8]
10) "protected": binary field that indicates whether the account has been set as protected.[9]

As for machine learning models, we tested a variety of algorithms readily available in the Python toolbox named *scikit-learn* (Pedregosa, et al., 2011).

In line with the considerations above, we considered the ability of the algorithms to deal with large datasets, which excluded some computationally more demanding algorithms (e.g., Support Vector Machines) and we benchmarked the following methods: Logistic Regression, Decision Trees, various ensemble methods (Random Forests, AdaBoost, ExtraTrees, etc.), *K*-nearest neighbors, Stochastic Gradient Descent, and finally two-layer neural networks.

For performance evaluation, we used two standard metrics commonly adopted in machine learning research, namely *accuracy* and *AUC-ROC* (Area Under the Receiver Operating Characteristic curve) (Bishop, 2016). Both scores range between zero and one, the larger the better, with one indicating perfect classification.

We set up a traditional supervised learning task, constituted of three phases, namely models' training, validation (a.k.a. performance evaluation), and finally, classification of the users in the Twitter French election dataset.

The first step required us to train each model with labeled examples of the two classes of users to detect (i.e., humans and bots). To this purpose, we used two datasets containing over five thousand of positive (bots) and negative (humans) examples of Twitter users in each category. The former training dataset is associated with *Botometer* (Varol, et al., 2017); the latter one is a labeled dataset provided by Cresci and collaborators (Cresci, 2017).

For performance evaluation, to calculate the accuracy and AUC-ROC scores of all models, we used the approach of *10-fold cross-validation*. This procedure splits the training data into ten equally-sized sets of data-points (preserving the balance of positive and negative data-points of the original dataset): one of these folds is hold out for validation (i.e., performance evaluation) and the remainder are used for training the models (the procedure is iterated 10 times, each holding out a different fold, and then averaging the accuracy and AUC-ROC scores obtained across the ten rounds of cross validation).

All models achieved very good performance, above 80% in both accuracy and AUC-ROC scores. The top three models in terms of performance were Random Forests (93% accuracy, 92% AUC-ROC), AdaBoost (92% accuracy and AUC-ROC), and Logistic Regression (92% accuracy, 89% AUC-ROC). The latter also was over one order of magnitude faster than nearly any other model (only Stochastic Gradient Descent was comparable in terms of speed but significantly worse in terms of performance).

For the reasons above, we decided to use Logistic Regression as reference model for bot detection purposes in the rest of this study. We retrained a full Logistic Regression model on the ten, simple metadata and activity features described above, using all the available labeled training data. Finally, we used it to classify all two million users in the Twitter French election dataset. An in-depth analysis of our findings follows.

# DATA ANALYSIS

## TIMELINE OF EVENTS AND ONLINE DISCUSSION DYNAMICS

We start by exploring the timeline of the general election-related discussion on Twitter. The broader discussion that we collected concerns the two candidates, Marine Le Pen and Emmanuel Macron, and spans the period from April 27 to May 7, 2017. The discussion revolves around the 23 key terms listed in **Table**

1 (Desigaud, et al., 2017; Howard, et al., 2017) and accounts for roughly 16.65 million tweets. **Figure 1** illustrates the timeline of the volume of such tweets, at the granularity of the minute, and with GMT-0 time zone alignment. Let us discuss first the dashed grey line (left axis): this shows the volume of generic election-related discussion. Note that the presidential election occurred on May 07, 2017. The discussion exhibits common circadian activity patterns and a slightly upwards trend in proximity to Election Day. Some spikes do occur, namely on May 3 and on May 7, 2017. Aside from the obvious uptake in discussion on Election Day (i.e., spike on May 7) driven by the offline events, the previous spike occurred at night time on May 3, 2017 again in response to an offline event, namely after the televised political debate that saw Le Pen facing Macron.[10] Otherwise, the number of tweets per minute averages between 300 and 1,500 during the day, and quickly approaches de facto zero overnight, consistently throughout the entire observation window.

**Figure 1** also illustrates with the purple solid line (right axis) the volume associated with MacronLeaks. One should first note that the volume is nearly an order of magnitude lesser than the general election-related discussion. However, the temporal pattern of this campaign is substantially different from the general conversation. First, the campaign is substantially silent for the entire period till early May. We can easily pinpoint to the inception of the campaign on Twitter, which occurs in the afternoon of April 30. After that, a surge in the volume of tweets, peaking at nearly 300 per minute, happens in the run up to Election Day, between May 5 and May 6, 2017. It has been already reported by prior research (Metaxas & Mustafaraj, 2012) that such disinformation campaigns peak between one and two days before elections. It is also worth noting that such a peak is nearly comparable in scale to the volume of the regular discussion, suggesting that for a brief interval of time (roughly 48 hours) the MacronLeaks disinformation campaign acquired significant collective attention, which in turn could have potentially had disastrous effects in terms of public opinion manipulation.

To understand the main topics of conversation and the main actors therein involved, in **Table 2** and **Table 3** we report the list of top 20 hashtags and mentioned users, respectively, with the associated number of tweets produced during our observation window. From **Table 2** we can observe that many terms that we used as keywords for data collection appear highly ranked in this list as well (cf. **Table 1**). Macron canalized by far the largest volume of tweets, with #Macron appearing in over 1.5 million tweets. Comparatively, the official hashtags of Marine Le Pen, namely #LePen, accrued less than one third of that. Macron's prominent presence in the discussion was due in part to organic attention, and in part because of disinformation: #MacronLeaks and #MacronGate, the two main hashtags related to the disinformation campaign, feature prominently in this top 20 list and appear cumulatively in about 350,000 tweets in our dataset. More details about this appear in the next section. Finally, the list demonstrates the high quality of our data collection strategy: all hashtags in the list are strictly related to the election conversation (we manually scrutinized up to the top 100 hashtags and results are consistent with very little noise added as we go down the ranking).

As far as the top mentioned users, the ranking is intuitively led by the two official accounts of the candidates, with Macron accruing nearly 30% more mentions than Le Pen. Several accounts follow, divided by a large gap. The list includes:

- Other prominent politicians such as Nicolas Dupont-Aignan, Florian Philippot, Jean-Luc Mélenchon, Marion Maréchal-Le Pen, and even US presidential candidate Hillary Clinton, as well as several official party Twitter accounts.

- Traditional news media accounts, including news channels like BFMTV, Quotidien (@qofficiel), as well as prominent journalists like Hugo Clément.
- American alt-right media personalities like InfoWars' editor Paul Joseph Watson (@prisonplanet), alt-right activist and *The Rebel*'s correspondent Jack Posobiec, as well as WikiLeaks. All these actors were prominently involved with sharing and discussing the MacronLeaks contents.

In the next two sections, we will focus on the MacronLeaks campaign and study social bot operations and characteristics; afterwards, we will describe the behaviors of human and bot users as well as their interactions. These analyses will shed light on the dynamics of the MacronLeaks disinformation operations.

## MACRONLEAKS BOTS AND THEIR CHARACTERISTICS

By using a Logistic Regression model trained on the ten metadata and activity features described above, we obtain very accurate user classification on the cross-validated tests (92% accuracy, 89% AUC-ROC). We adopted this model to detect all bots and separate them from human users in our dataset. In the following, however, we will focus exclusively on the MacronLeaks corpus, the subset of our Twitter data that contains nearly 350,000 tweets posted by nearly one hundred thousand distinct users.[11]

Out of 99,378 users involved in MacronLeaks, our model classified **18,324** of them as social bots, and the remainder of 81,054 as human users. The fraction of social bots amounts for about 18% of the total users involved in the campaign, which is extremely consistent with results from previous studies – e.g., our analysis of the 2016 U.S. Presidential election uncovered that roughly 15% of the users involved in the Twitter conversation were bots, and accounted for about 20% of the total tweets (Bessi & Ferrara, 2016).

To provide a rigorous and thorough assessment of the quality of the bot detection results, next we provide first some examples of the results generated by our bot detection model (cf. **Table 4**), and then a broader statistical characterization of the distinctive features exhibited by the bots as opposed to human users.

**Table 4** shows the list of the top 15 Twitter accounts detected as social bots in the MacronLeaks corpus by our Logistic Regression model. They are sorted by the number of tweets they posted during our observation window. We manually investigated the status of these accounts at the time of our investigation (end of May 2017): The column "*Status*" reports whether the accounts are still active, or otherwise they have been deleted by the owner, or suspended or "quarantined" by Twitter.[12] Whether an account has been deleted, suspended, or quarantined, it is a strong indicator that the account has been involved in activities in violation of Twitter's Terms of Service: for example social bots, when detected by the platform's algorithms, get systematically shut down; users who share information incompatible with Twitter's ToS can get reported by others, then scrutinized by Twitter's anti-abuse team, and suspended or quarantined if found in violation.

Remarkably, among the top 15 social bots detected by our framework, 4 accounts have been so far deleted, 7 have been suspended, and 2 have been quarantined by Twitter. Two accounts are still active, and they may be the result of misclassification. Overall, this example of manual verification suggests that we obtained 13 correct bots out of 15 detected, an accuracy of nearly 87% which is compatible with the cross-validation benchmark (92% accuracy, 89% AUC-ROC). Further inspection of highly-ranked accounts is consistent with this accuracy performance.

In general, we have already extensively observed in our prior work how detecting bots "in the wild" is a much more challenging task than traditional machine learning "exercises" where performance is measured on validation test sets for which labels are known (Ferrara, et al., 2016; Varol, et al., 2017). This discrepancy happens for a couple of reasons, most prominently because oftentimes bots used during traditional training-validation benchmarks are of the same or similar types, while in the wild (i.e., in the real Twitter world) one can expect that thousands of variants of social bots may exist. Our detection framework seems to perform well in real-world detection. We hypothesize that this happens for two reasons. First, the framework benefits from its simplicity: Logistic Regression is a very simple linear model, and we limit the model to learn only over ten user metadata and activity feature, so overfitting issues are limited). Second, to train our models we used a mixture of various types of bots provided by multiple studies (Varol, et al., 2017; Cresci, et al., 2017).

Let us provide some characteristics of the 13 hand-verified bots in **Table 4**:

- All accounts exhibit a disproportionate number of tweets containing the MacronLeaks keywords, generated over the first week of May 2017.
- 12 out of 13 bots (all except @Yhesum) appear to have a very limited number of followers, suggesting that their creators did not emphasize the importance of the social network dimension, which has proven to be a central component of success and influence of bot operations in prior studies (Bessi & Ferrara, 2016; Ferrara, et al., 2016). The only exception (@Yhesum) instead appears to adopt the well-known automatic reciprocal follow-back strategy (i.e., retaining as friends who follow back an automatically-initiated bot followership), as suggested by the very large and balanced number of friends and followers (nearly ten thousand).
- Two suspicious groups of bot accounts with similar names appear:
    - A first class of bots is named *2020 (where * is a randomly generated name, e.g., *gunbuster2020*, *dixneuf2020*, etc.). Four such "2020" bots appear in the top 15 and they all exhibit similar activity statistics. Further manual scrutiny revealed the existence of three additional "2020" bots in the MacronLeaks corpus, which were less active but still identical in behavioral patterns to the four in the top 15 list.
    - A second class of bots is named *_1337 (where * is a randomly generated name, e.g., Geoff_1337 and jerry_1337). Further scrutiny revealed the existing of five other "_1337" less active bots. It is worth noting that *1337* spells *leet,* which stands for "elite" and represents an alternative alphabet that is primarily used by hacking internet communities.[13]

In the **Discussion and Conclusions** section, we will present some additional insights about our findings related to the activity and characteristics of some of the bots we discovered. We will show some evidence in support of the hypothesis that online markets of reusable political disinformation bots may exist:[14] some of the bot accounts we uncovered were created at the beginning of November 2016, shortly before the 2016 U.S. Presidential election, and used only for a week to support of alt-right narratives; then they "went dark", showing no activity till early May, in support of alt-right agenda and the MacronLeaks disinformation campaign in the context of the 2017 French Presidential election. Evidence of such devices being in other contexts, e.g., in case of crises, has been already reported (Starbird, et al., 2014; Nied, et al., 2017).

We enrich our analysis by providing a statistical characterization of the discovered bots, and we contrast it to the accounts identified as human users by our framework. To this purpose, in **Figure 2** and **Figure 3** we show two boxplot distributions of the activity and metadata features respectively of the human users and the social bots present in the MacronLeaks Twitter corpus. We focus only on the five discrete features, namely number of tweets (*statuses_count*), number of followers (*followers_count*), number of friends (*friends_count*), number of favorites (*favourites_count*), and number of times listed (*listed_count*); thus, we exclude the other five binary features, i.e., *default_profile*, *geo_enabled*, *profile_use_background_image*, *verified*, and *protected*.

Some strong statistical differences clearly emerge: Along all dimensions, the social bots involved in MacronLeaks appear as less active than the human users. Pairwise comparisons between bots and humans of the distributions of these five features yield additional insights:

- Bots posted on average 2.86 MacronLeaks-related tweets ($\sigma = 10.3$), while humans posted 3.81 ($\sigma = 9.68$). A t-test and a Mann-Whitney test yield respectively p-values of $10^{-18}$ and $3^{-80}$.
- Bots obtained on average 1,382 followers ($\sigma = 22,282$), while humans have 2,510 ($\sigma = 28,542$). T-test and Mann-Whitney test score respectively p-values of $3^{-05}$ and zero.
- Bots friended on average 1,058 users ($\sigma = 12,190$), while humans have on average 1,403 friends ($\sigma = 3,656$). T-test and Mann-Whitney yield p-values of $2^{-04}$ and zero.
- Bots favorited on average 228 tweets ($\sigma = 924$), while humans have on average 13,774 favorites ($\sigma = 27,001$). Both t-test and Mann-Whitney p-values are zero.
- Finally, bots have been listed on average on 7.42 lists ($\sigma = 90.3$) and humans instead on 77.64 ($\sigma = 560.2$). T-test yields a p-value of 2-62 and Mann-Whitney a p-value of zero.

All p-values indicate that the pairs of distributions differ statistically very significantly. It is worth noting that all distributions are broad and skewed, exhibiting a power-law like behavior. For a visual example, refer to **Figure 6**, which will be discussed in detail in the next section.

These results warrant further discussion. Recent literature reported extensively on traditional types of social bots whose automated activity can be easily detected because it yields very high volumes of tweets, especially retweets, due to incessant posting and re-sharing operations (Ferrara, et al., 2016; Bessi & Ferrara, 2016). These are historically considered effective strategies as they aim at flooding the platform with campaign-related contents and canalize collective attention. However, in the case of MacronLeaks the historic trend appears to be reversed: Bots seem to try "fly under the radars", posting comparable or less content than humans in the same conversation. We hypothesize that this can be a strategy to avoid detection and suspension from the platform.

We finally investigate whether any correlation between user activity and metadata emerge that could indicate further anomalous behavioral patterns in bots as opposed to human users. To this aim, in **Figure 4** we show the feature Pearson correlation heat maps for human users (top), and social bots (bottom figure) – exclusively for those accounts involved in the MacronLeaks disinformation campaign. Two very different pictures emerge: for human users (N = 81,054) no strong pairwise feature correlations occur, as measure by the Pearson correlation coefficient, except for a strong correlation ($\rho = 0.79$) between the number of followers a human user has and the number of times it appears in public lists (cf. **Figure 4, top**). These two features are intuitively correlated because more influential users with more followers are added to public lists more frequently than less influential users with fewer followers.

A very different picture emerges from the pairwise feature correlation of bot accounts (N = 18,324). The strongest correlation (cf. **Figure 4, bottom**) appears between the number of friends and the number of times these bots appear in lists (Pearson coefficient $\rho = 0.81$): this indicates that the more users a bot follows the more often it appears in public lists of other users, suggesting a self-promotion mechanism commonly adopted by bots. The second strongest correlation emerges between the numbers of followers and friends (Pearson coefficient $\rho = 0.65$), indicating that on average the bots in the MacronLeaks corpus engage in automatic, reciprocal follower-friend links (one example of such behavior is the bot @Yhesum of **Table 4**). To complete the cluster of correlated pairs of features, number of friends and favorites correlate with a coefficient $\rho = 0.61$, and number of favorites and list appearance correlate with a coefficient of $\rho = 0.41$, both indicating strong correlated patterns of systematic behavior in the bots.

Overall, our analyses highlighted examples of bots involved in MacronLeaks as well as provided a statistical characterization of their activity and metadata features used by our framework to detect them. In the next section, we will explore more in detail the behavior of both bots and human users involved in MacronLeaks as well as their interactions.

## HUMAN AND BOT BEHAVIOR AND HUMAN-BOT INTERACTIONS

Like in the previous section, we here first provide a temporal characterization of the Twitter activity, this time specifically related to MacronLeaks, for both bot and human accounts. In **Figure 5**, we show the timeline of the volume of tweets generated (granularity of one minute), respectively by human users (dashed grey line) and social bots (solid purple line), between April 27, 2017 and May 7, 2017, and related to MacronLeaks. The amount of activity is substantially very close to zero until May 5, 2017, in line with the first coordination efforts as well as the information leaks spurred from other social platforms like 4chan.org, as discussed in the introduction. Spikes in bot-generated content often appear to slightly precede spikes in human posts, suggesting that bots can trigger cascades of disinformation. At peak, the volume of bot-generated tweets is comparable with the that of human-generated ones (the plot, however, does not differentiate between humans' retweets of bot-generated content, as opposed to tweets organically generated by human users).

Let us investigate human-bot interactions further, and specifically determine what are the characteristics of the bots that are more frequently retweeted. To this aim, in **Table 5** we report a list of ten social bots frequently retweeted by human users. Intuitively, some accounts overlap with the list of most active bots reported in **Table 4** – e.g., the @Yhesum bot, which we discussed extensively before, accrued many retweets. All the accounts in **Table 5** are suspended, deleted, or quarantined by Twitter at the time of this writing (early June 2017): this corroborates the suspect that the accounts were involved in the incriminated activities. The examples of **Table 5** warrant further discussion, which will allow us to dig more into the details of the strategies adopted by the bots and their effectiveness.

For each of the example bot accounts appearing in **Table 5**, we calculated the number of followers these accounts had prior to MacronLeaks (more precisely, the first time they appear in our dataset) as opposed to at the end of the observation window (i.e., relative to their last tweet in our dataset). What emerges is that six out of ten bots accrued a significant and large number of followers during the MacronLeaks disinformation campaign. A few of these example accounts had zero or nearly no followers when they first appear in the conversation, and manual verification indicates that the accounts were explicitly created on purpose to disseminate MacronLeaks tweets. It is worth noting that being active in such a type of disinformation campaigns appears to be an effective strategy to accrue visibility: several accounts obtained thousands of followers – in our example bots of **Table 5**, we notice increases up to nearly fifteen thousand followers accumulated in the span of less than one week. Another interesting fact that is worth noting is that the number of retweets received by the bots does not seem to be strongly correlated with the total amount of MacronLeaks tweets posted by each bot. The effectiveness of these bots, if measured in terms of number of accrued retweets, appear to be independent of their sheer activity: this suggests that other factors, such as the position of the bots in the social network, the different types of messages they posted and their narratives and language, as well as the type of human targets they aim at influencing, are all ingredients that potentially play a role in bots' strategies.

To dig further into the type of users involved in MacronLeaks and contrast that to the users participating exclusively to the general election-related discussion, we extracted the most frequent words occurring in the tweets produced in these two corpora. In **Table 6** we show as example the top 16 most frequent token terms (i.e., individual words obtained after removing stop words and other commonly occurring terms) occurring in the tweets of the two corpora. Excluding the names of the two candidates that are prominently featured in both corpora, by contrasting the two columns stark differences emerge: on the left side of **Table 6**, which represents general-election frequent terms, we observe several French words that strongly relate to the upcoming voting, e.g., *voter* (to cast one's vote, in French), *vote* ("to vote", in French as in English), *fait* (fact, or event, in French), *faire* ("to do", "to make", "to take", etc.), *tout* (all), *Français* (French), *débat* (debate), etc. On the right side of **Table 6** we see no such a language characterization, which strongly suggests the fact that the MacronLeaks disinformation campaign was limited mostly to an English-speaking audience, and failed to percolate in the French-speaking Twitter community. The most frequent words speak along the directions of the main narrative that was tailored around the alleged illicit activities and tax frauds of the now-president Macron. Interestingly, the word *campaign* itself is prominently featured in fourth position with over twenty thousand distinct occurrences, suggesting the nature of the operation itself.

To corroborate this hypothesis, we extracted the most frequent terms appearing in the self-reported profile description that Twitter users can decide to include in their accounts. By using the same filtering criteria, in **Table 7** we report as example the top 16 word tokens ranked by frequency of appearance in the Twitter profiles of users involved respectively in MacronLeaks (right column) as opposed to exclusively in the general election-related discussion (left column). **Table 7** provides again a staggeringly different picture: most words occurring in the profiles of users not involved in MacronLeaks campaigns are in French, while emblematically the top two key terms of MacronLeaks' users are *MAGA* ("Make America Great Again", the motto of US President Trump), as well as *Trump* itself. While the left column is suggestive of terms that individuals interested in French politics would likely adopt to describe themselves – *politique* (politics), *patriote* (patriot), *Français* (French), *contre* (against), *gauche* (left), *droite* (right), etc. – the right side features similar key terms in English (conservative, patriot, god, pro-, American, America, Christian, politics, anti-, life, supporter, country, etc.), which would clearly characterize a right-leaning English-speaking, American audience. This suggests that the largest majority of users involved in MacronLeaks had prior interests in American politics, in right-wing narratives, and in alt-right political agenda. Manual inspection of many of these accounts confirmed this hypothesis; it also further revealed that some accounts labeled as bots involved in MacronLeaks were also active in the alt-right campaigns leading to the 2016 U.S. Presidential election and were labeled as bots in our previous study as well (Bessi & Ferrara, 2016).

We next analyze the type of information sources that the users, respectively human and bots, most frequently referred to in their tweets. To this purpose, in **Table 8** we show for illustrative example the list of the top 10 URLs that have been tweeted within the general election discussion. We will consider this as a baseline for comparison and contrast it to the top tweeted URLs shown in **Table 9**, and discussed later. Our extensive analysis highlighted that most URLs appearing in the general election-related discussion are of the following three types:

1. Pointers to tweets of either presidential candidate, (Macron and Le Pen), or to tweets of other prominent politicians (e.g., Nicolas Dupont-Aignan);
2. Pointers to articles published in established news media (Le Figaro, The Guardian, Le Monde);

3. Pointers to influential users on external media channels (e.g., YouTube, etc.), as well as journalists and other influential users on Twitter.

Let us look at the MacronLeaks corpus next. **Table 9** shows the list of top 10 URLs tweeted within the MacronLeaks campaign. In stark contrast with the general election-related discussion, the list of top tweeted URLs contains a mix of:

1. Links to hyper-partisan news outlets: the top tweeted URL points to a story appeared on *The Gateway Pundit* (an alt-right news media) claiming that leaked documents were original and credible. Other highly-tweeted stories appear again on *The Gateway Pundit* as well as on other alt-right websites like *Zero Hedge* and *GotNews* and support the allegations of Macron's financial frauds and illicit activities.
2. Links to leaked data dumps: the second most tweeted URL is the link to the *archive.is* file that contains the data of the leaked documents; this file has been extensively circulated on other platforms like 4chan.org and was shared thousands of times on Twitter as well; and, finally,
3. URLs pointing to fake news websites: some blogposts (e.g., on *blogspot.dk*) qualify as fake news as they seem to be designed exclusively for financial profit, rather than for political reasons, e.g., in the case of alt-right news media sites like *Breitbart News, Infowars, The Gateway Pundit,* etc. Pointers to fake news stories proliferate even further down the list of highly-retweeted URLs.s

We conclude this analysis section by providing a statistical characterization of the corpus' features. In **Figure 6** we show two sets of probability distributions, relative to the MacronLeaks tweet corpus (cf., top figure), which contains 347 thousand tweets, and an equal-sized random sample of tweets related to the general French election discussion posted during the same period (cf., bottom figure).

In this order, each plot shows:

      *A.* The distribution of the total number of tweets posted by each user in the given corpus;
      *B.* The distribution of number of total word tokens in the given tweet corpus;
      *C.* The distribution of number of total word tokens in the user profiles' descriptions;
      *D.* The distribution of the number of tweets' languages;
      *E.* The distribution of the number of distinct hashtags;
      *F.* The distribution of the number of distinct user mentions; and finally,
      *G.* The distribution of the number of distinct URLs, appearing in the corpus.

Before contrasting the results for the two corpora, let us provide an intuitive explanation of these features.

We considered the number of tweets as an indicator of volume of activity in a discussion: we hypothesize that in a bot-driven conversation, highly active accounts (due to automation) will contribute giving this distribution a fatter tail than in an organic discussion.

The features *B* and *C* warrant an explanation on their construction: for each corpus (i.e., MacronLeaks and general discussion, respectively), we calculated the frequency of appearance of each word in the tweets (*B)* as well as in the user profile descriptions (*C*). Word distributions enjoy some characteristic distributions, e.g., the Zipf law that suggests that the ranked word frequency should obey a Zipfian distribution (Adamic, 2000). Our intuition is that the heavier the fat tail of such a distribution, the less diverse a tweet corpus is (respectively, the user profiles appearing in it), because a fatter tail indicates the appearance of some extremely popular words (in either the tweets or the user profile descriptions).

Feature *D* simply indicates how many different languages appear in a corpus, and the frequency of tweets in that language. A more diverse discussion should encompass more languages. We hypothesize that the ranking of the languages is also instrumental to determine the audience mainly involved in that discussion.

The last three features, *E, F,* and *G,* indicate the number of distinct hashtags, user mentions, and URLs appearing in the given corpus. More diverse conversations will have lighter tails than discussions dominated by the pervasive frequency of some of these entities.

Let us now discuss the findings that emerge from **Figure 6**, following the order of features presented above: for what concerns the distribution of the total number of tweets posted by each user in the two corpora (cf., distribution *A, i.e., the solid blue line*), we can see that in MacronLeaks the tail of the distributions reaches roughly $10^3$, indicating the presence of some extremely prolific accounts (either human or bots); in contrast, such distribution characterizing the general discussion exhibits a smaller slope and thus peaks at around $10^2$, suggesting that only a minority of users posted more than one hundred tweets in the entire corpus. This is in line with the intuition that bot-driven discussion exhibit a fatter tail in the characteristic activity-related distributions (e.g., number of posted tweets per user).

Distributions *B (i.e., the dashed red line)* and *C (i.e., dashed orange line)* are very similar one another in both corpora. These distributions reflect the frequency of occurrence of word terms, whose generation and interpretation have been extensively discussed above in reference to **Table 6** and **Table 7**, which report examples of word token appearing in the tweets and in the users' profile descriptions, respectively. Once again, the slopes of the curves relative to MacronLeaks is larger than that of the general discussion, suggesting the presence of more prominent words that appear more frequently in the tweets as well as in the profiles of users in MacronLeaks. This is consistent with the examples shown in **Table 6** and **Table 7**, which nicely illustrate two byproduct effects of this phenomenon:

1. The most frequent word in both rankings appears about 25% more in MacronLeaks than in the general discussion: "Macron" appears 133K times in MacronLeaks, but only 109K times in the general discussion; similarly, "MAGA" (top ranked in **Table 7: right column**) appears nearly 18K times in MacronLeaks' user descriptions, as opposed to "France" (top ranked in **Table 7: left column**) that appears about 13K times in the profiles of general discussion users.
2. As one scrolls down the ranking, given a position in the ranking, terms in MacronLeaks corpus tent to have higher frequencies of occurrence of equally ranked terms in the general election. For example, the tenth terms in **Table 6** are *fait*, which appears 11,023 times in the general election-related discussion, and *fakes* that appears 12,404 in MacronLeaks. Similarly, the twentieth terms in **Table 7** are *droite,* which appears 3,871 times in the general discussion, and *country* that appears 5,272 times in MacronLeaks. The same applies to nearly all entries in the rest of the rankings.

These results corroborate our intuition that fatter distribution tails suggest less diversity in the tweet corpus, or in the profiles of the users therein appearing. In MacronLeaks this indicates the emergence of popular words strongly evocative of the corpus' narrative and prominently featured by a relatively larger number of users with respect to the general election-related discussion.

Yet in **Figure 6**, Distribution *D* provides a view into the diversity of languages feature in each corpus. Intuitively, a more diverse conversation that is more inclusive of a variety of groups, as well as that attracts international attention, would feature a larger number of languages. MacronLeaks once again shows a fatter tail suggesting that the conversation is less diverse in terms of languages given that many tweets appears in one extremely popular language. Further analysis reveals an interesting fact: in MacronLeaks, most tweets

(N=177,695) are in English, while French is only second (N=135,397); in stark difference, in the equal-sized random sample of general conversation, French emerges by far as the most prominent language (N=242,422), while English comes second with a large gap (N=73,409). This suggest that the main participants to MacronLeaks were not in the French-speaking community, but rather in the English-speaking American user base. This may be one explanation associated with the scarce success of the disinformation campaign to affect French voters.

Our analysis concludes with a summary of interpretations of the last three distributions, *E, F,* and *G,* which indicate the number of distinct hashtags, user mentions, and URLs respectively. For each of these distributions it is possible to observe systematically roughly one order of magnitude difference in favor of the MacronLeaks ones, which exhibit fatter tails indicating less diverse corpora whose tweets are dominated by fewer, more popular hashtags, mentions, and URLs, if contrasted with the general discussion. If one considers URLs, as an example, the results are reflected in **Table 8** and **Table 9** and a similar analysis regarding ranking and frequencies of the terms can be done as for what concerned distribution *B* and **Table 6** and **Table 7**: the top URLs in MacronLeaks are systematically more frequent than equally-ranked URLs in the general discussion.

# DISCUSSION AND CONCLUSIONS

In this paper, we provided an extensive statistical analysis of the MacronLeaks disinformation campaign that occurred in the run up to the 2017 French presidential election. Using a mix of state-of-the-art machine learning techniques and cognitive heuristics for bot detection, in combination with data-driven insights, and considering a reconstruction of events as they unfolded, we uncovered some characteristics of the disinformation campaign, its main drivers, and its human audience. We contrasted these results against the general election-related conversation that we used as baseline to pinpoint to differences and anomalies.

Our results highlighted a few interesting phenomena: first, we advanced the hypothesis that a black market of reusable political disinformation bots may exist. Similar suggestions have been advanced by other studies (Starbird, et al., 2014; Nied, et al., 2017); however, our work is the first to identify the presence of bots that existed during the 2016 U.S. Presidential election campaign period to support alt-right narratives, went dark after November 8, 2016, and came back into use in the run up days to the 2017 French presidential election.

In conclusion, our findings also demonstrated that the prior interest of users engaged in MacronLeaks may be revealing of the reasons of scarce success of the campaign at affecting the French vote outcome: most of the audience of MacronLeaks campaign was the English-speaking American alt-right community, rather than French users; this is in stark contrast with the baseline general conversation, which involved systematically and significantly more French users (thus, likely French voters), which exhibited a clear trend in favor of supporting now-president Emmanuel Macron.

In the future, we will try to draw similarities and differences in the context of computational political propaganda phenomena, focusing on a variety of elections and politics-related contemporary events to understand how online social media can be manipulated, and what are the quantifiable consequences of successful such attempts.

# NOTES

1. While the exact dynamics of events in the context of the 2016 U.S. election at the time of this writing (June 2017) still remain unclear, and are subject of ongoing federal investigations, sufficient evidence has been mounting to support the idea that foreign governments, as well as organizations with vested interests, may have meddled with the election process, either via influence campaigns, or by means of spear phishing attacks aimed at stealing private information of voters either for retaliation and blackmailing purposes.
2. See: https://www.cnet.com/news/macron-french-presidential-campaign-says-it-was-hacked/
3. See: http://www.bbc.com/news/blogs-trending-39845105
4. See: http://www.thedailybeast.com/the-twitter-bots-who-tried-to-steal-france
5. See: https://www.buzzfeed.com/craigsilverman/partisan-fb-pages-analysis
6. *Botometer* is publicly available at: https://botometer.iuni.iu.edu/
7. We here use a nomenclature of tweets' features consistent with that of the Twitter API; See official documentation as reference: https://dev.twitter.com/overview/api/tweets
8. Twitter account verification process described at: https://support.twitter.com/articles/119135
9. About public and protected tweets: https://support.twitter.com/articles/14016
10. Timeline of events leading to the election: http://www.businessinsider.com/france-is-having-one-of-its-strangest-presidential-elections-timeline-2017-5
11. A broader analysis of social bot operations in the entire French election dataset is left for future work. This is mostly because a thorough validation of the bot classification results requires a significant amount of manual investigation for quality assurance. Therefore, scaling up the effort to very large corpora (in our case, 2 million users overall) requires significant additional resources.
12. Twitter can take two different countermeasures toward suspicious accounts: the more radical solution is an account suspension, which yields the deletion of the account and all tweets ever posted by that user. A softer solution is the so-called "quarantine": the given account is labeled as suspicious and users on the platform must go through an extra step of verification to access their content. The owner of a quarantined account can also try appeal to Twitter to get reinstated.
13. See: https://en.wikipedia.org/wiki/Leet
14. See: http://www.businessinsider.com/twitter-bots-and-fake-accounts-2013-11

# ABOUT THE AUTHOR

Dr. **Emilio Ferrara** is Research Assistant Professor at the University of Southern California, Research Leader at the USC Information Sciences Institute, and Principal Investigator of the USC Machine Intelligence and Data Science (MINDS). His research focuses on characterizing information diffusion in online social networks, detecting and preventing abuse in such systems. He was named *2015 IBM Watson Big Data Influencer*, he is recipient of the *2016 Complex System Society Junior Scientific Award*, and he received the *2016 DARPA Young Faculty Award*. **E-mail**: emiliofe@usc.edu

# ACKNOWLEDGEMENTS

This work is supported in part by the Air Force Office of Scientific Research. The funder had no role in study design, data collection and analysis, manuscript preparation, or decision to publish.

# TABLES

| | | | |
|---|---|---|---|
| France2017 | LePen | JeVoteMacron | EnMarche |
| Marine2017 | Le Pen | JeVote | MacronPresident |
| AuNomDuPeuple | MarineLePen | Presidentielle2017 | #France |
| FrenchElection | FrenchPresidentialElection | ElectionFracaise | @MLP_officiel |
| FrenchElections | JeChoisisMarine | JamaisMacron | @EmmanuelMacron |
| Macron | JeVoteMarine | Macron2017 | |

**Table 1**. List of 23 keywords whose streams we continuously collected from April 27, 2017 to May 7, 2017.

| Hashtag | No. Tweets | Hashtag (cont.) | No. Tweets (cont.) |
|---|---|---|---|
| #Macron | 1,521,425 | #fn | 127,336 |
| #Presidentielle2017 | 652,563 | #JeVote | 126,075 |
| #LePen | 447,365 | #Marine2017 | 119,274 |
| #France | 378,234 | #debat2017 | 118,069 |
| #2017LeDebat | 273,304 | #2017LeDébat | 98,862 |
| #MacronLeaks | 255,491 | #MacronGate | 93,235 |
| #MarineLePen | 184,442 | #JeVoteMacron | 83,081 |
| #Whirlpool | 170,347 | #MacronPresident | 72,184 |
| #EnMarche | 128,613 | #JamaisMacron | 72,140 |
| #FrenchElection | 127,705 | #Elysee2017 | 59,421 |

**Table 2**. List of top 20 hashtags and associated number of tweets produced between April 27, 2017 and May 7, 2017 (election day).

| Mention | No. Tweets | Mention (cont.) | No. Tweets (cont.) |
| --- | --- | --- | --- |
| @emmanuelmacron | 1,284,627 | @hugoclement | 66,001 |
| @mlp_officiel | 981,219 | @enmarchefr | 62,468 |
| @bfmtv | 213,421 | @hillaryclinton | 61,982 |
| @dupontaignan | 178,995 | @marion_m_le_pen | 61,651 |
| @prisonplanet | 136,874 | @sofiakkar | 58,074 |
| @qofficiel | 112,423 | @v_of_europe | 56,998 |
| @f_philippot | 95,696 | @jeunesmacron | 54,927 |
| @jlmelenchon | 89,334 | @flwjrm | 50,982 |
| @jackposobiec | 76,419 | @youtube | 50,631 |
| @wikileaks | 69,718 | @tf1lejt | 50,628 |

**Table 3**. List of top 20 mentioned users, and associated number of tweets produced between April 27, 2017 and May 7, 2017 (election day).

| Username | Tweets | Followers | Friends | Favorites | Listed | Status[*] |
|---|---|---|---|---|---|---|
| Vote__Marine | 737 | 9 | 61 | 85 | 0 | DELETED |
| DonTreadOnMemes | 427 | 24 | 35 | 371 | 1 | DELETED |
| BrandonJBarber | 328 | 4 | 135 | 0 | 1 | DELETED |
| Geoff_1337 | 286 | 14 | 13 | 19 | 0 | SUSPENDED |
| N********* | 252 | 31 | 99 | 270 | 4 | Active |
| 07_mai_2017 | 225 | 169 | 231 | 0 | 3 | DELETED |
| gunbuster2020 | 221 | 10 | 87 | 574 | 0 | SUSPENDED |
| dixneuf2020 | 213 | 15 | 81 | 570 | 0 | SUSPENDED |
| bluecanti2020 | 204 | 10 | 88 | 575 | 0 | SUSPENDED |
| shogouki2020 | 203 | 13 | 86 | 573 | 0 | SUSPENDED |
| D********* | 199 | 2,515 | 1,676 | 77 | 29 | Active |
| Yhesum | 191 | 9,476 | 9,599 | 223 | 96 | SUSPENDED |
| jerry_1337 | 152 | 5 | 12 | 0 | 0 | SUSPENDED |
| Subocean | 102 | 26 | 87 | 9 | 3 | QUARANTINED |
| protegerlepeupl | 92 | 1 | 62 | 60 | 0 | QUARANTINED |

**Table 4**. List of the top 15 Twitter accounts detected as social bots by our algorithms and sorted according to the number of MacronLeaks posts they tweeted between April 27, 2017 and May 7, 2017 (election day). [*] The column "*Status*" reports whether, as of the date of this writing (end of May 2017), the account is active or is being suspended, deleted, or "quarantined" by Twitter. Out of the top 15 social bots detected by our framework, 4 accounts have been so far deleted, 7 have been suspended, and 2 have been quarantined by Twitter. Two accounts are still active (their usernames are redacted to preserve user privacy).

| Username | Retweeted | Tweets | Followers before | Followers after | Status[*] |
|---|---|---|---|---|---|
| Yhesum | 291 | 786 | 21 | 9,476 | SUSPENDED |
| lou_justine92 | 119 | 213 | 219 | 2,297 | DELETED |
| trololo451 | 93 | 94 | 7 | 10 | DELETED |
| doubtallthought | 43 | 431 | 377 | 1,833 | QUARANTINED |
| chadashworth | 43 | 5 | 14,833 | 15,205 | DELETED |
| jewishhotjean | 39 | 422 | 46 | 14,033 | DELETED |
| sebasqien | 38 | 447 | 31 | 6,633 | DELETED |
| blufor2 | 30 | 3 | 0 | 1 | DELETED |
| _loup_gar | 29 | 65 | 2 | 769 | DELETED |
| lerenardfrance | 25 | 22 | 6 | 16 | DELETED |

**Table 5.** Ten examples of social bots frequently retweeted by human users. [*]All these accounts are suspended, deleted, or quarantined by Twitter at the time of this writing (early June 2017).

| Generic election-related discussion | | MacronLeaks campaign | |
| --- | --- | --- | --- |
| **Token** | **Frequency** | **Token** | **Frequency** |
| Macron | 109,734 | Macron | 133,708 |
| Pen | 80,789 | French | 28,668 |
| Marine | 43,101 | France | 27,280 |
| France | 24,971 | campaign | 22,585 |
| Emmanuel | 15,793 | emails | 19,206 |
| French | 13,463 | Pen | 15,463 |
| voter | 13,094 | WikiLeaks | 14,493 |
| vote | 11,394 | documents | 14,269 |
| plus | 11,204 | tax | 12,591 |
| fait | 11,023 | fakes | 12,404 |
| faire | 9,237 | election | 12,159 |
| LePen | 8,717 | media | 10,053 |
| election | 7,311 | leaked | 9,643 |
| Français | 6,535 | vote | 9,641 |
| tout | 6,497 | discovered | 8,927 |
| débat | 5,930 | evasion | 8,854 |

**Table 6**. Top 16 most frequent token terms occurring in the tweets of users respectively active in the general election-related discussion or in the MacroLeaks campaign.

| Generic election-related discussion | | MacronLeaks campaign | |
|---|---|---|---|
| **Token** | **Frequency** | **Token** | **Frequency** |
| France | 13,204 | MAGA | 17,853 |
| politique | 7,731 | Trump | 17,796 |
| love | 6,241 | France | 15,437 |
| patriote | 5,578 | love | 12,345 |
| marine2017 | 5,576 | conservative | 10,657 |
| Trump | 5,116 | marine2017 | 8,503 |
| vie | 5,104 | patriot | 8,280 |
| monde | 5,041 | marine | 7,386 |
| tweets | 4,874 | god | 7,297 |
| fan | 4,713 | proud | 6,931 |
| Français | 4,667 | pro | 6,880 |
| news | 4,488 | American | 6,777 |
| contre | 4,359 | America | 6,774 |
| anti | 4,301 | Christian | 6,110 |
| gauche | 4,247 | politics | 5,625 |
| Macron | 4,187 | anti | 5,603 |
| Marine | 4,105 | fan | 5,579 |
| EnMarche | 3,997 | life | 5,414 |
| ans | 3,944 | supporter | 5,332 |
| droite | 3,871 | country | 5,272 |

**Table 7**. Top 20 most frequent token terms appearing in the user account self-reported descriptions of users active respectively in the general election discussion, or in the MacronLeaks campaign.

| URL | Frequency |
|---|---|
| https://twitter.com/mlp_officiel/status/5968249560764416 | 1,064 |
| https://twitter.com/mlp_officiel/status/858932738942021632 | 470 |
| https://twitter.com/madamefigaro/status/860134337806835712 | 417 |
| https://twitter.com/dupontaignan/status/858033334521409536 | 411 |
| https://www.youtube.com/watch?v=D6H0cjIN4gw | 306 |
| https://twitter.com/MLP_officiel/status/858932738942021632 | 259 |
| https://www.theguardian.com/world/2017/may/04/barack-obama-backs-macron-in-last-minute-election-intervention | 243 |
| https://twitter.com/damienrieu/status/858265285991780352 | 213 |
| https://twitter.com/anneapplebaum/status/860872956498644992 | 191 |
| http://www.lemonde.fr/les-decodeurs/article/2017/05/03/des-intox-du-debat-entre-emmanuel-macron-et-marine-le-pen-verifiees_5121846_4355770.html | 171 |

**Table 8**. Top 10 URLs tweeted within the general election discussion (baseline). Most URLs are pointers to tweets of either presidential candidate, other prominent politicians (e.g., Nicolas Dupont-Aignan), or established news media (Le Figaro, The Guardian, Le Monde).

| URL | Frequency |
| --- | --- |
| http://www.thegatewaypundit.com/2017/05/breaking-wikileaks-confirms-leaked-macron-campaign-emails-authentic-macronleaks/ | 3,018 |
| http://archive.is/eQtrm | 2,766 |
| http://www.zerohedge.com/news/2017-04-25/meet-real-emmanuel-macron-consummate-banker-puppet-bizarre-elitist-creation | 2,077 |
| http://gotnews.com/busted-macronleaks-show-feminist-hypocrite-emmanuelmacron-pays-female-campaign-workers-26-less-men/ | 1,943 |
| https://www.pscp.tv/w/a9nszjF4ZUtXeEdWcnJhalB8MUJkeFl2bWtwa0RLWI2X5c-aHtYkbfBlqL9jLCqUAewt8H54OrqhULHt16_h | 1,780 |
| http://gotnews.com/emmanuel-macrons-tax-evasion-documents-real/ | 1,718 |
| http://disobedientmedia.com/new-leak-reveals-emails-documents-from-macron-and-affiliated-staff-members/ | 1,542 |
| http://www.thegatewaypundit.com/2017/05/breaking-macron-busted-lied-tax-evasion-4chan-pol-posts-images-macrons-off-shore-bank-account/ | 673 |
| http://disobedientmedia.com/macron-denies-authenticity-of-leak-french-prosecutors-open-probe/ | 641 |
| https://diversitymachtfrei.blogspot.dk/2017/05/macron-leaks-contain-secret-plans-for.html | 635 |

**Table 9**. Top 10 URLs tweeted within the MacronLeaks campaign. The list contains a mix of links to hyper-partisan news outlets (*The Gateway Pundit, Zero Hedge, GotNews),* leaked data dumps, and fake news websites.

# FIGURES

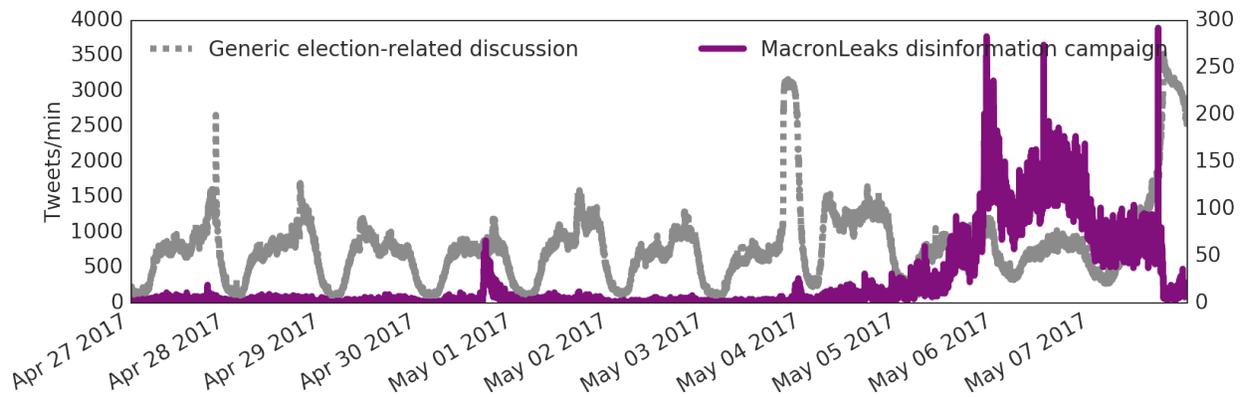

**Figure 1**. Timeline of the volume of tweets generated every minute during our observation period (April 27, 2017 through May 7, 2017). The purple solid line (right axis) shows the volume associated with MacronLeaks, while the dashed grey line (left axis) shows the volume of generic election-related discussion. The presidential election occurred on May 07, 2017.

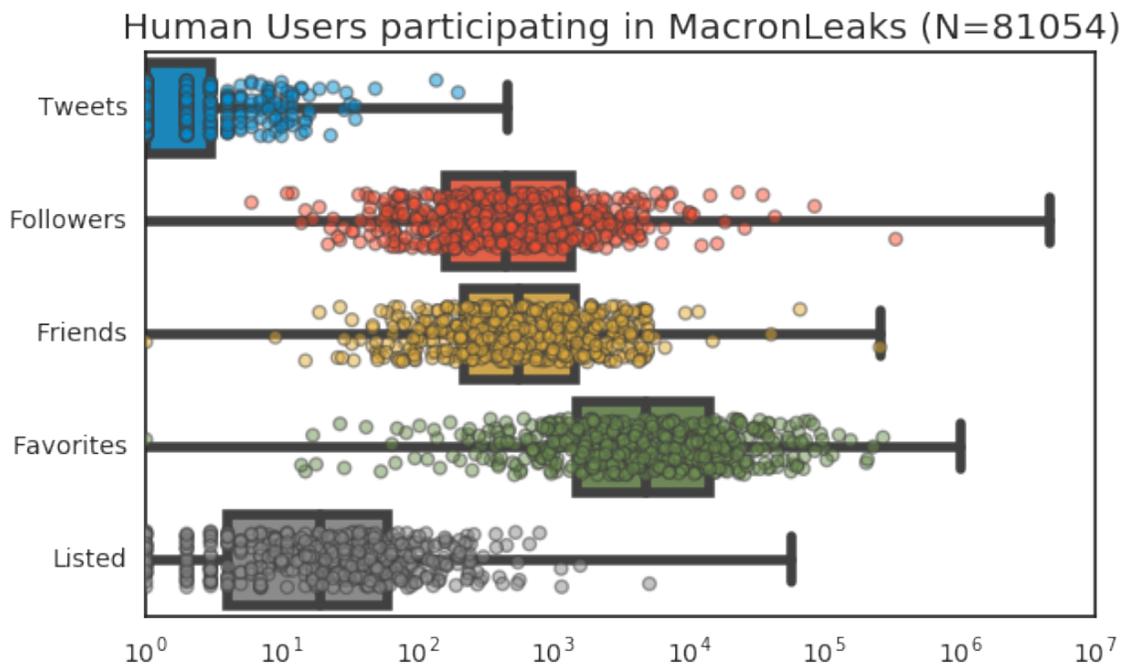

**Figure 2**. Boxplot distribution of the metadata features of to the human users involved in the disinformation campaigns associated with MacronLeaks.

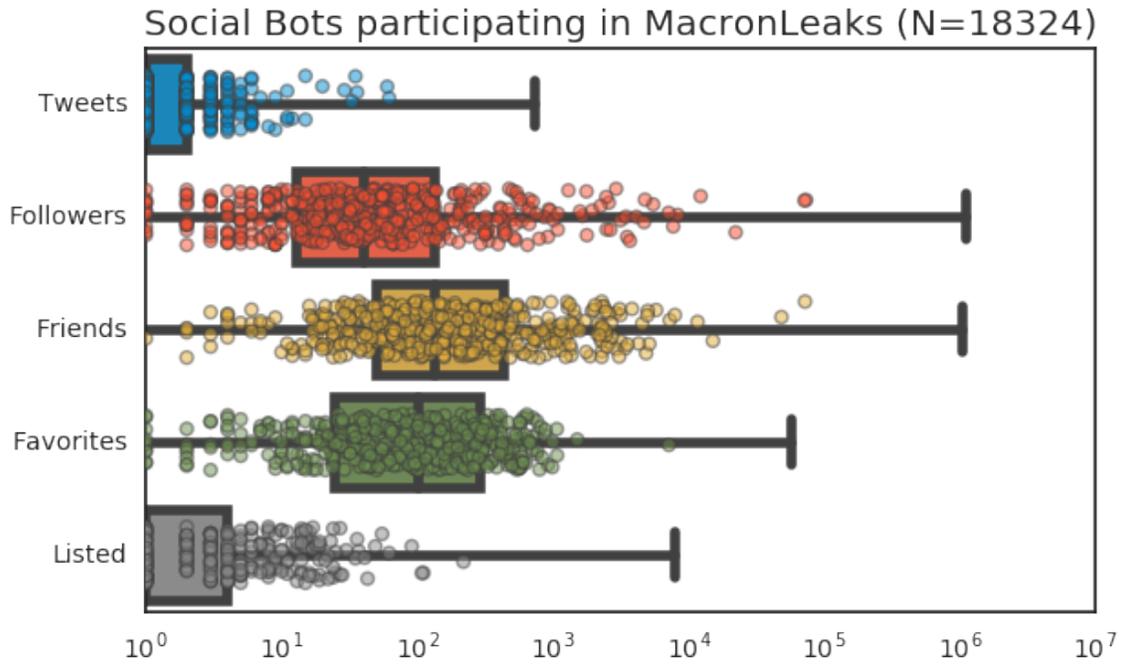

**Figure 3**. Boxplot distribution of the metadata features of the social bots detected by our framework and associated with the disinformation campaigns related to MacronLeaks.

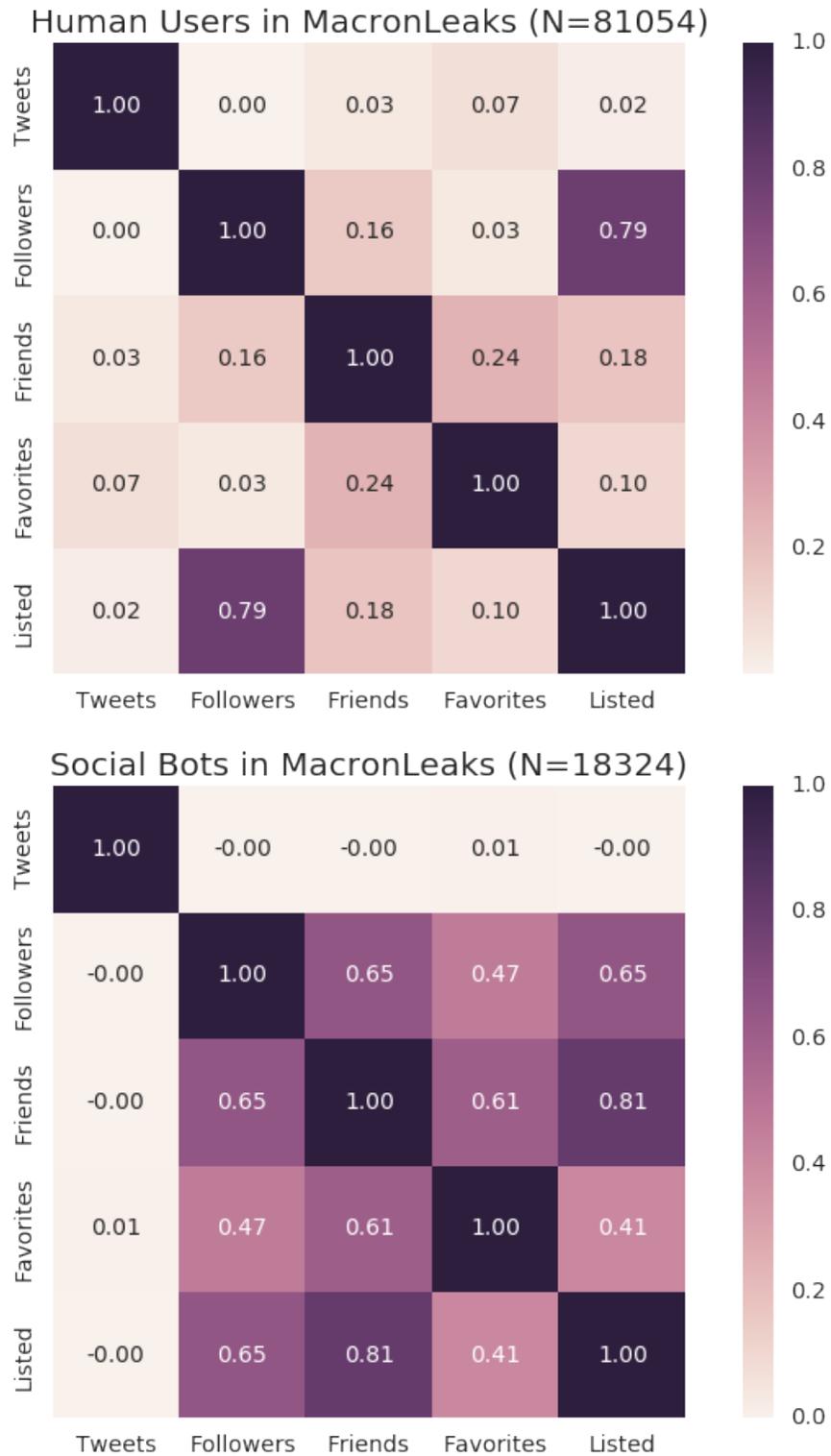

**Figure 4**. Feature correlation heat maps for human users' metadata (**TOP**), and social bots' features (**BOTTOM**) involved in the disinformation campaigns associated with MacronLeaks.

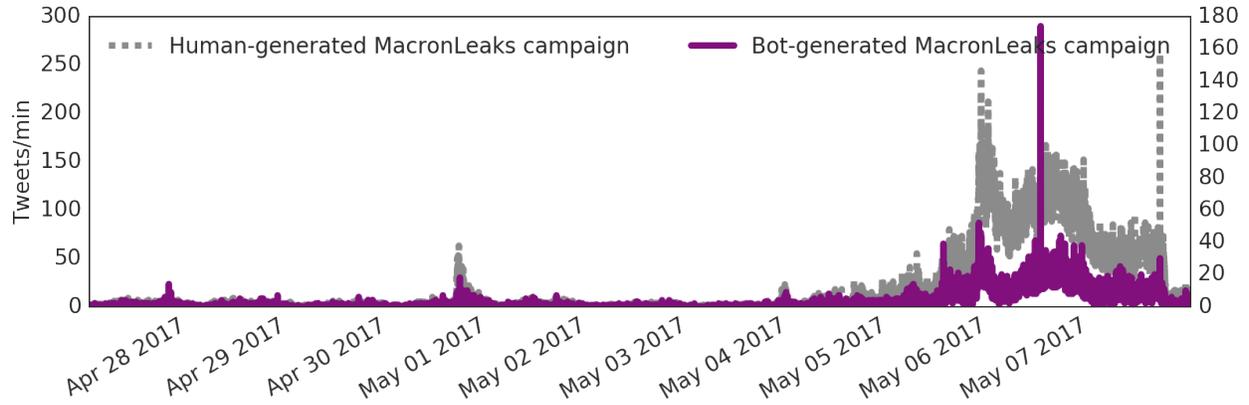

**Figure 5.** Timeline of the volume of tweets generated every minute, respectively by human users (dashed grey line) and social bots (solid purple line), between April 27, 2017 and May 7, 2017, and related to MacronLeaks. Spikes in bot-generated content often slightly precedes spikes in human posts, suggesting that bots can trigger cascades of disinformation.

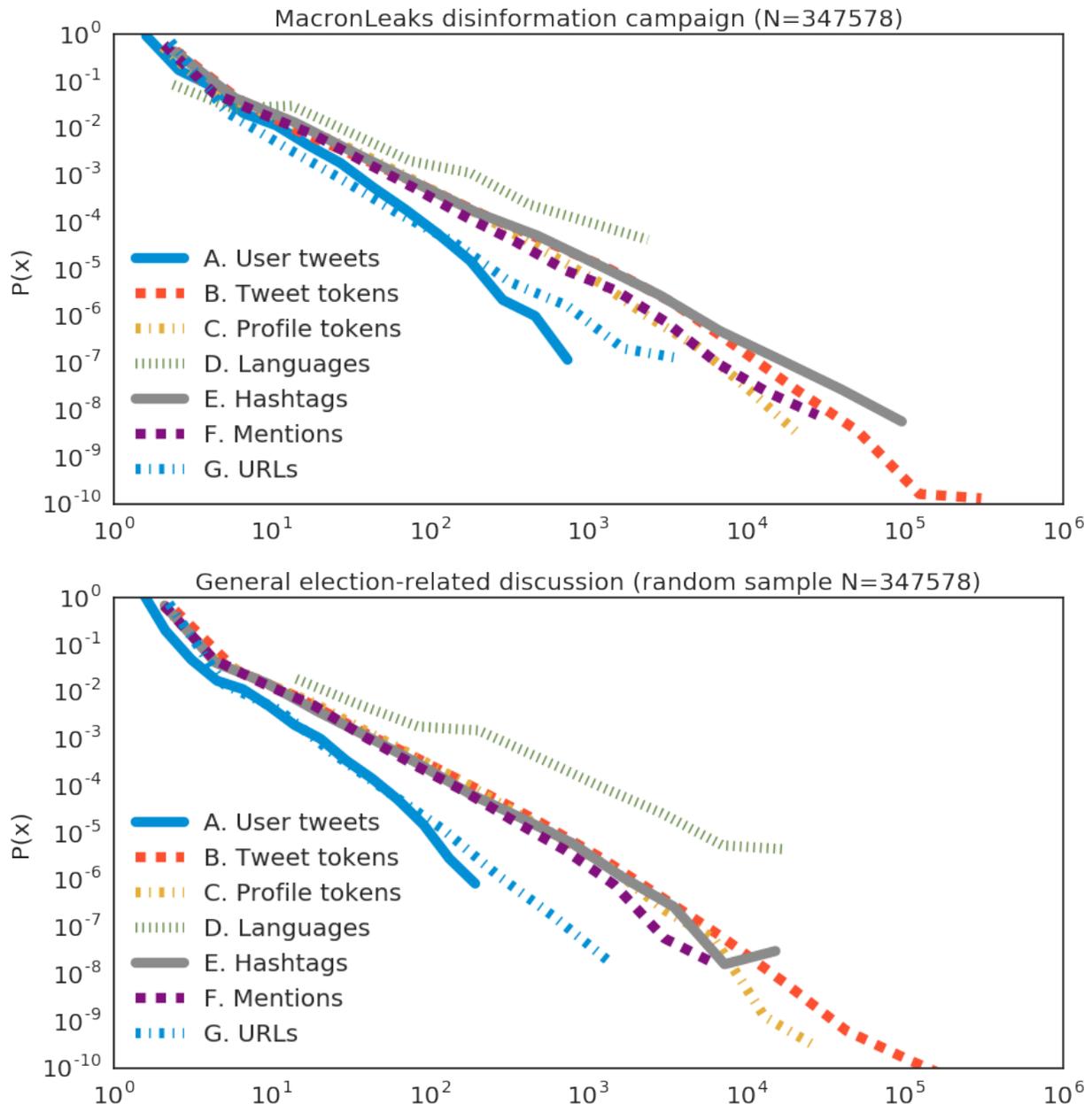

**Figure 6**. **TOP**: distributions of statistics calculated on the MacronLeaks tweet corpus (347*K* tweets); **BOTTOM**: statistics calculated on an equal-sized random sample of tweets related to the French election posted during the same period (April 27, 2017 through May 7, 2017). The plots show, in this order: *(A)* the distribution of the number of tweets posted by each user; *(B)* the distribution of number of total word tokens in the tweet corpus, as well as *(C)* in the user profiles' descriptions; *(D)* the distribution of the number of tweets' languages; the distribution of the number of distinct *(E)* hashtags, *(F)* user mentions, and finally *(G)* URLs, appearing in the corpus.